\documentclass[12pt]{iopart}
\usepackage{iopams}
\usepackage{graphicx,wrapfig}

\newcommand{\rShock}{R_{\mbox{\tiny Sh}}}
\newcommand{\rPNS}{R_{\mbox{\tiny PNS}}}
\newcommand{\model}[1]{\mbox{B}{#1}}
\newcommand{\timeAverage}[3]{\langle{#1}\rangle_{#2\hspace{0.05cm}\mbox{\tiny s}}^{#3\hspace{0.05cm}\mbox{\tiny s}}}

\newcommand{\vect}[1]{{\boldsymbol{#1}}}
\newcommand{\pderiv}[2]{\frac{{\partial #1}}{{\partial #2}}}
\newcommand{\f}[2]{\frac{#1}{#2}}

\newcommand{\curl}[1]{\nabla\times{#1}}

\begin{document}

\title[Turbulence and magnetic field amplification from spiral SASI modes]{Turbulence and magnetic field amplification from spiral SASI modes in core-collapse supernovae}

\author{E Endeve$^{1}$, C Y Cardall$^{2,3}$, R D Budiardja$^{2,3,4}$, A Mezzacappa$^{1,2,3}$, and J M Blondin$^{5}$}

\address{$^1$ Computer Science and Mathematics Division, Oak Ridge National Laboratory (ORNL), Oak Ridge, Tennessee, USA}
\address{$^2$Physics Division, ORNL, Oak Ridge, TN 37831, USA}
\address{$^3$Department of Physics and Astronomy, University of Tennessee, Knoxville, TN 37996, USA}
\address{$^4$Joint Institute for Heavy Ion Research, ORNL, Oak Ridge, TN 37831, USA}
\address{$^5$Physics Department, North Carolina State University, Raleigh, NC 27695-8202, USA}
\ead{endevee@ornl.gov}

\begin{abstract}
  The stationary accretion shock instability (SASI) plays a central role in modern simulations of the explosion phase of core-collapse supernovae (CCSNe).  
  It may be key to realizing neutrino powered explosions, and possibly links birth properties of pulsars (e.g., kick, spin, and magnetic field) to supernova dynamics.  
  Using high-resolution magnetohydrodynamic simulations, we study the development of turbulence, and subsequent amplification of magnetic fields in a simplified model of the post-bounce core-collapse supernova environment.  
  Turbulence develops from secondary instabilities induced by the SASI.  
  Our simulations suggest that the development of turbulence plays an important role for the subsequent evolution of the SASI.  
  The turbulence also acts to amplify weak magnetic fields via a small-scale dynamo.  
\end{abstract}
\submitto{\PS Proceedings of Turbulent Mixing and Beyond (TMB-2011).}

\maketitle
 
\section{Introduction}

The core-collapse supernova (CCSN) blast is initiated when the central density of a collapsing massive star ($\gtrsim 8 M_\odot$) exceeds nuclear density.  
At this point the repulsive short-range nuclear force stiffens the stellar core's equation of state, and the resulting core bounce launches a compression wave that turns into a shock when it reaches the supersonically collapsing outer core.  
As the roughly spherical shock propagates radially outward through the collapsing layers of the core, it loses energy through neutrino emission and dissociation of heavy nuclei, eventually stalling to form an accretion shock some 100-200~km from the center of the star.  
It is expected that the shock is revived within a second or so, enabling it to disrupt the star's outer layers and give rise to the supernova.  

In the current paradigm, initiated by Bethe \& Wilson \cite{betheWilson_1985}, the shock is revived by energy deposition in the gain region below the shock from the intense flux of neutrinos streaming out of the forming proto-neutron star (PNS).  
However, sophisticated spherically symmetric supernova models fail to reproduce the explosion \cite{ramppJanka_2000, liebendorfer_etal_2001, thompson_etal_2003}, except in the case of the lightest supernova progenitors with O-Ne-Mg cores \cite{kitaura_etal_2006}.  
Results from multidimensional multiphysics simulations (e.g., \cite{burrows_etal_2006,bruenn_etal_2009,marekJanka_2009,suwa_etal_2010,muller_etal_2012}) are encouraging, but discrepant results from independent research groups, in part due to the physics included in the models and the approximations made, remain to be fully investigated.  
It appears clear, however, that a detailed understanding of the explosion mechanism will likely involve the confluence of multi-frequency (or multi-frequency {\it and} multi-angle) neutrino transport, nuclear physics (equation of state and reaction kinetics), rotation, and (magneto)hydrodynamic instabilities \cite{mezzacappa_2005, woosleyJanka_2005,janka_etal_2007}.  

The stationary accretion shock instability (SASI \cite{blondin_etal_2003}) is central to modern simulations of neutrino-powered supernova explosions \cite{bruenn_etal_2006,buras_etal_2006,bruenn_etal_2009,marekJanka_2009,suwa_etal_2010,muller_etal_2012}.  
The dominant SASI modes are often characterized in terms of low-order spherical harmonics $Y_{\ell}^{m}$ ($\ell=1$, $m=0,\pm1$) \cite{blondinMezzacappa_2006,blondinShaw_2007}.  
In the early phase, the SASI manifests itself through global oscillations of the stalled shock, while in the later stages it results in post-shock turbulence (driven by SASI-induced downdrafts from the shock) and an overall expansion of the shocked cavity.  
The instability may facilitate the explosion by improving the conditions for energy deposition by neutrinos (e.g., \cite{marekJanka_2009}).  
In terms of physical fidelity, the most detailed simulations have so far been carried out in two spatial dimensions (2D; with axial symmetry imposed), but early simulations in three spatial dimensions (3D) seem to confirm the importance of the SASI (e.g., \cite{fryerYoung_2007,takiwaki_etal_2011}).  
Aside from the explosion mechanism itself, the SASI may also connect supernova dynamics with birth properties of neutron stars (e.g., natal kick \cite{scheck_etal_2004}, spin \cite{blondinMezzacappa_2007}, and magnetic field \cite{endeve_etal_2010}).  

To further explore the development of the SASI in 3D, we present results from high-resolution magnetohydrodynamic (MHD) simulations.  
We adopt the adiabatic model used by Blondin et al. \cite{blondin_etal_2003,blondinMezzacappa_2007} and study the development and impact of SASI-induced turbulence by perturbing the non-rotating, spherically symmetric initial condition.  
The so-called spiral SASI mode ($\ell=1,|m|=1$) \cite{blondinMezzacappa_2007,blondinShaw_2007,fernandez_2010} develops favorably in our simulations.  
Vigorous turbulence develops from secondary instabilities in the shear layer separating two counterrotating streams induced by the spiral SASI mode.  
The turbulence feeds on the power in the low-order SASI modes, and our simulations suggest that the SASI saturates nonlinearly from the development of turbulence (as was proposed by Guilet et al. \cite{guilet_etal_2010}).  
The SASI-driven turbulence is essentially non-helical (e.g., \cite{brandenburg_etal_1996,haugen_etal_2004}), and results in an efficient small-scale dynamo.  
The initially weak magnetic fields grow exponentially by turbulent stretching until the magnetic energy becomes comparable (locally) to the kinetic energy of the turbulent flows.  
The potential relevance of our simulation results to multiphysics simulations of CCSNe and neutron star birth properties is discussed.  

\section{Numerical simulations}

To study the development of SASI-driven turbulence and associated magnetic field amplification, we solve the ideal MHD equations (e.g., \cite{landauLifshitz_1960}) for mass density, fluid velocity, internal energy density, and magnetic field ($\rho$, $\vect{u}$, $e$, and $\vect{B}$, respectively).  
We adopt the ideal gas equation of state, which relates the pressure to the internal energy density $P=(\gamma-1)e$, where $\gamma$ is the adiabatic index.  
The gravitational potential around the central compact object (the PNS) is approximated with the point-mass formula $\Phi=-GM/r$, where $G$ is Newton's constant, $M$ the mass of the central object, and $r$ the radial distance from the center.  
A time-explicit, second-order HLL-type finite volume scheme is used for the time-integration of the ideal MHD equations (see \cite{endeve_etal_2010,endeve_etal_2012a,endeve_etal_2012b} for details).  

Our numerical simulations are initiated with the adiabatic setup used by Blondin et al. and Blondin \& Mezzacappa \cite{blondin_etal_2003,blondinMezzacappa_2007} (see also \cite{endeve_etal_2010}), which resembles the early stalled shock phase in a core-collapse supernova: a spherical, stationary accretion shock is placed at a radius $\rShock=200$~km.  
A supersonic flow is nearly free-falling towards the shock for $r>\rShock$.  
Between the shock and the PNS, the flow settles subsonically in nearly hydrostatic equilibrium.  
Matter is allowed to flow through an inner boundary placed at $\rPNS=40$~km.  
The adiabatic index is set to $\gamma=4/3$ (similar to the degenerate conditions in the collapsing core).  
The mass of the central object $M=1.2~M_{\odot}$ and the accretion rate ahead of the shock $\dot{M}=0.36~M_{\odot}\mbox{ s}^{-1}$ are held fixed during the simulations.  
A radial (split monopole) magnetic field is superimposed on the initial condition: $B_{r}=\mbox{sign}(\cos\vartheta)\,B_{0}\,(\rPNS/r)^{2}$, where $B_{0}$ is the magnetic field strength at the surface of the PNS.  
We initiate the SASI with small ($\sim1\%$) random pressure perturbations in the post-shock flow.  

We have carried out simulations on a three-dimensional Cartesian grid with resolution $\Delta l/\rPNS\approx0.029$, and using up to $1280^{3}$ cells ($\Delta l$ is the width of a cell).  
Here we present simulations where we have varied the initial magnetic field: $B_{0}=1\times10^{10}$~G (model $\model{10}$ or ``weak-field model'') and $B_{0}=1\times10^{13}$~G (model $\model{13}$ or ``strong-field model'').  
The initial field can be considered weak in both models in the sense that the magnetic energy density is small compared to the kinetic and internal energy densities.  

\section{Turbulence from spiral SASI modes}
 
\begin{figure}
  \begin{center}
    \includegraphics[angle=00,height=2.44in,width=6.3in]{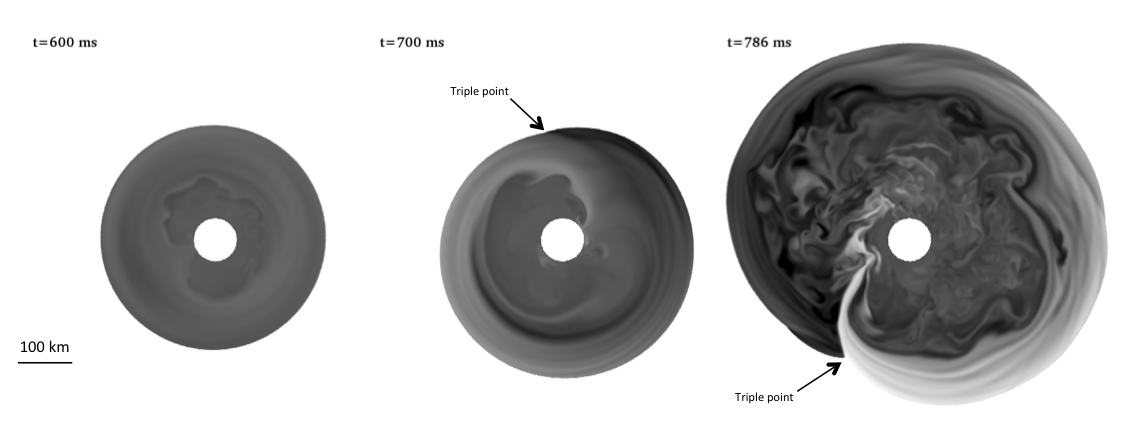}
    \caption{Select snapshots showing the development of the spiral SASI mode, and subsequent development of turbulence below the shock.  The polytropic constant $\kappa=P/\rho^{\gamma}$ (a proxy for fluid entropy) is displayed for times $t=600$~ms, $700$~ms, and $786$~ms (left, middle and right, respectively).  Darker regions are associated with higher entropy.  \label{fig:entropy}}
  \end{center}
\end{figure}

The spiral SASI mode emerges favorably from our numerical simulations.  
\Fref{fig:entropy} displays snapshots of the fluid entropy and provides a visual impression of the development of the spiral SASI mode and turbulence in the strong-field model (the weak-field model evolves similarly).  
Early on ($t=600$~ms), the SASI manifests itself by small-amplitude global oscillations of the still spherical shocked cavity.  
In terms of an expansion in spherical harmonics $Y_{\ell}^{m}$, the shock oscillations are often characterized as sloshing modes ($\ell=1, m=0$) or spiral modes ($\ell=1, m=\pm1$).  
However, sloshing and spiral modes are related: spiral SASI modes can be viewed as a superposition of sloshing modes out of phase, and thus may be a more general outcome of perturbing the spherically symmetric initial condition (non-radially) \cite{blondinShaw_2007,fernandez_2010,foglizzo_etal_2012}.  
The SASI results in angular momentum redistribution and counterrotating flows inside the shock.  

The spiral SASI mode leads to strong velocity shear in the layer separating the two counterrotating streams in the post-shock flow.  
With growing amplitude, the spiral SASI mode eventually leads to the formation of a shock triple point (cf. \fref{fig:entropy}).  
The shock triple point forms when a shock wave from a steepening pressure wave, propagating on the inside the accretion shock, connects with the accretion shock itself in the vertex separating the counterrotating post-shock flows and the pre-shock accretion flow \cite{blondinMezzacappa_2007}.  
The shock triple point (in reality a line segment extending across the accretion shock surface) has just formed at $t=700$~ms and is located in the top portion of the middle panel in \fref{fig:entropy}.  
It propagates in the counterclockwise direction.  
Lower-entropy material ahead of the triple point (flowing in the clockwise direction) is diverted down towards the PNS, while higher-entropy material trails the triple point.  

Vigorous turbulence emerges as a result of the nonlinear evolution of the SASI.  
The shear layer extending from the triple point down towards the PNS is vulnerable to the development of secondary fluid instabilities (e.g., the Kelvin-Helmholtz instability) and turbulence.  
In the right panel in \fref{fig:entropy} ($t=786$~ms), the shock triple point has completed one and a half revolutions about the PNS, and the flow appears turbulent in a large fraction of the shocked volume.  
(The low-entropy stream towards the PNS is supersonic at this time.)  
The turbulent nature of the post-shock flow can also be seen in the vorticity field $\vect{\omega}=\curl{\vect{u}}$, which exhibits high spatial and temporal intermittency---in part due to the formation of vortex tube structures---in the late stages of the evolution \cite{endeve_etal_2010,endeve_etal_2012a}.  

\begin{figure}
  \begin{center}
    \includegraphics[angle=00,height=4.0in,width=4.0in]{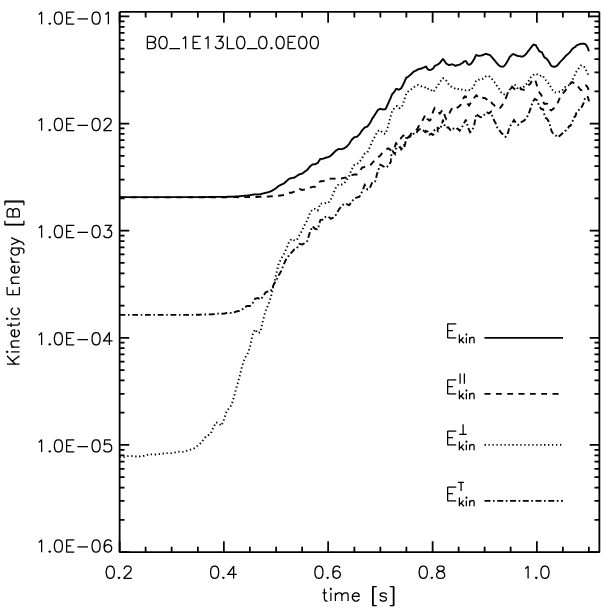}
    \caption{Evolution of the kinetic energy below the shock:  total kinetic energy $E_{\mbox{\tiny kin}}=\f{1}{2}\int_{V_{\mbox{\tiny Sh}}}\rho u^{2}\,dV$, kinetic energy associated with radial flow $E_{\mbox{\tiny kin}}^{\parallel}=\f{1}{2}\int_{V_{\mbox{\tiny Sh}}}\rho u_{r}^{2}\,dV$ and non-radial flow $E_{\mbox{\tiny kin}}^{\perp}=\f{1}{2}\int_{V_{\mbox{\tiny Sh}}}\rho (u_{\vartheta}^{2}+u_{\varphi}^{2})\,dV$, and kinetic energy due to small-scale ``turbulent" flows $E_{\mbox{\tiny kin}}^{\mbox{\tiny T}}=\int_{k^{\mbox{\tiny T}}}^{\infty} \widehat{e}_{\mbox{\tiny kin}}(k)\,dk$, where $\widehat{e}_{\mbox{\tiny kin}}$ is the spectral kinetic energy density (cf. \eref{eq:spectralDensity}).  In computing $E_{\mbox{\tiny kin}}^{\mbox{\tiny T}}$, we have set $k^{\mbox{\tiny T}}=0.1$~km$^{-1}$ (i.e., $\lambda^{\mbox{\tiny T}}=2\pi/k^{\mbox{\tiny T}}\approx63$~km).  The diamonds on the solid line correspond to times displayed in \fref{fig:entropy} (only the earliest three) and \fref{fig:kineticAndEnstrophySpectra}.  ($1~\mbox{Bethe [B]}\equiv10^{51}$~erg.) \label{fig:kineticEnergyVersusTime}}
  \end{center}
\end{figure}

The kinetic energy below the shock grows rapidly with time during the ramp-up of the SASI.  
In \fref{fig:kineticEnergyVersusTime}, we plot the total kinetic energy (solid line) in $V_{\mbox{\tiny Sh}}$---the volume bounded by the surface of the shock and the surface of the PNS.  
We also plot the kinetic energy associated with radial flow $E_{\mbox{\tiny kin}}^{\parallel}$, non-radial flow $E_{\mbox{\tiny kin}}^{\perp}$, and ``turbulent" flow $E_{\mbox{\tiny kin}}^{\mbox{\tiny T}}$ (dashed, dotted, and dash-dotted lines, respectively; see caption for definitions).  

\begin{figure}
  \begin{center}
    \includegraphics[angle=00,height=3.0in,width=3.0in]{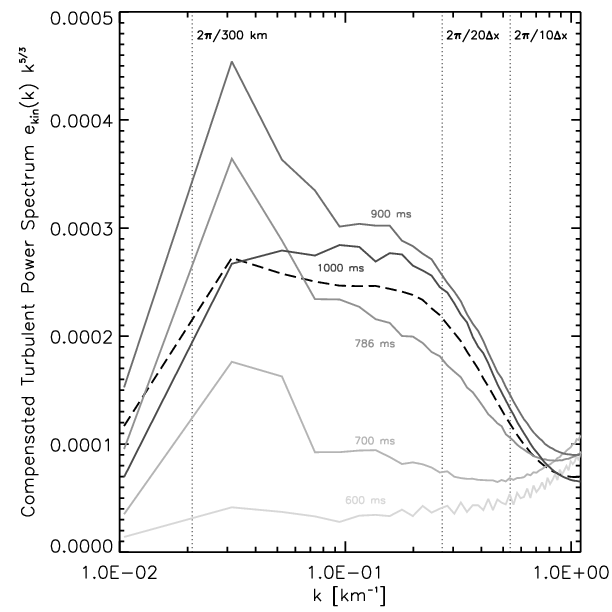}
    \includegraphics[angle=00,height=3.0in,width=3.0in]{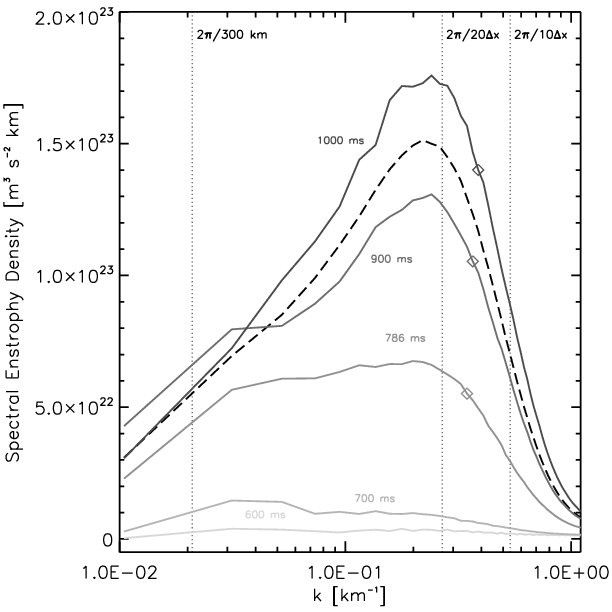}
    \caption{Evolution of Fourier spectra during SASI development in the strong-field simulation.  {\it Left panel:} compensated spectral specific kinetic energy density $\widehat{e}_{\vect{u}}\times k^{5/3}$.  {\it Right panel:} Spectral enstrophy density $\widehat{e}_{\vect{\omega}}$.  In both panels, dashed lines represent time averaged spectra (averaged over the time period extending from $t=1000$~ms to $t=1100$~ms).  The dotted vertical reference lines denote spatial scales corresponding to $300$~km, $20\,\Delta l$, and $10\,\Delta l$.  \label{fig:kineticAndEnstrophySpectra}}
  \end{center}
\end{figure}

Saturation of the kinetic energy below the shock coincides with the development of a significant turbulent component in the post-shock flow.  
The kinetic energy beneath the shock is initially about $2\times10^{-3}$~B, and the flow is essentially radial until $t\approx400$~ms.  
The kinetic energy begins to grow during the initial ramp-up phase of the SASI, which starts around 400~ms, and is due to the rapidly growing non-radial component.  
For $t=600$~ms, $E_{\mbox{\tiny kin}}^{\parallel}$ is still larger than $E_{\mbox{\tiny kin}}^{\perp}$.  
However, the non-radial kinetic energy exceeds the radial kinetic energy when $t=700$~ms, and the two components remain similar in magnitude for the remainder of the run.  
The kinetic energy growth slows down considerably around $t=786$~ms (cf. \fref{fig:entropy}), but it continues to grow throughout the nonlinear phase, with variability on a shorter timescale superimposed.  
The turbulent kinetic energy also grows rapidly during the ramp-up of the SASI, and becomes comparable (within a factor of two) to the radial kinetic energy.  
When averaged over the time interval extending from 900~ms to 1100~ms, the total kinetic energy below the shock is $\timeAverage{E_{\mbox{\tiny kin}}}{0.9}{1.1}=0.044$~B.  
Similarly, we find $\timeAverage{E_{\mbox{\tiny kin}}^{\parallel}}{0.9}{1.1}=0.019$~B, $\timeAverage{E_{\mbox{\tiny kin}}^{\perp}}{0.9}{1.1}=0.025$~B, and $\timeAverage{E_{\mbox{\tiny kin}}^{\mbox{\tiny T}}}{0.9}{1.1}=0.011$~B.  
Thus, about $25\%$ of the post-shock kinetic energy is associated with turbulence (this number, of course, depends on the value adopted for $k^{\mbox{\tiny T}}$ in the definition of $E_{\mbox{\tiny kin}}^{\mbox{\tiny T}}$, but we consider $k^{\mbox{\tiny T}}=0.1$ to be somewhat conservative, and the turbulent kinetic energy is possibly even larger; cf. \fref{fig:kineticAndEnstrophySpectra}).  
Moreover, the turbulent rms velocity $u_{\mbox{\tiny rms}}^{\mbox{\tiny T}}=\sqrt{2E_{\mbox{\tiny kin}}^{\mbox{\tiny T}}/M_{\mbox{\tiny Sh}}}$ is about $5000$~km~s$^{-1}$ during the saturated state ($M_{\mbox{\tiny Sh}}$ is the mass within $V_{\mbox{\tiny Sh}}$).  

Additional, quantitative insight into the development of turbulence from the SASI is gained by inspecting the Fourier spectra of select quantities from the simulations.  
From the Fourier transform of a vector field $\vect{f}$, 
\begin{equation}
  \widehat{\vect{f}}(\vect{k})
  =\f{1}{V_{L}}\int_{V_{L}}\vect{f}(\vect{x})\times\exp{(i\vect{k}\cdot\vect{x})}\,dV_{L}, 
  \label{eq:fourierTransform}
\end{equation}
where $V_{L}$ is the volume of the computational box, we obtain the spectral density per $k$-space shell
\begin{equation}
  \widehat{e}(k)=\f{1}{2}\int_{k\mbox{\tiny-Shell}}|\widehat{\vect{f}}|^{2}k^{2}\,d\Omega_{k}, 
  \label{eq:spectralDensity}
\end{equation}
where the magnitude of the wave vector (wavenumber) is $k=|\vect{k}|$, and $d\Omega_{k}$ is a solid angle element in Fourier space.  
When the spectral density is integrated over $k$-space, the result equals the real space integral of the square of the corresponding real space quantity; i.e., 
$\int_{k_{\mbox{\tiny min}}}^{k_{\mbox{\tiny max}}}\widehat{e}\,dk=\f{1}{2}\int_{V_{L}}|\vect{f}|^{2}\,dV$ ($k_{\mbox{\tiny min}}=2\pi/L$ and $k_{\mbox{\tiny max}}=2\pi/\Delta l$ are given by the size of the computational box $L$ and the grid spacing $\Delta l$, respectively).  
In particular, we obtain the spectral kinetic energy density $\widehat{e}_{\mbox{\tiny kin}}$ by setting $\vect{f}=\sqrt{\rho}\vect{u}$ in \eref{eq:fourierTransform}.  
In a similar manner, we obtain the spectral specific kinetic energy density $\widehat{e}_{\vect{u}}$, the spectral enstrophy density $\widehat{e}_{\vect{\omega}}$, and the spectral magnetic energy density $\widehat{e}_{\mbox{\tiny mag}}$ by setting $\vect{f}$ in \eref{eq:fourierTransform} to $\vect{u}$, $\vect{\omega}$, and $\vect{B}$, respectively.  
(When computing the spectra, we exclude the pre-shock flow, which is not of interest here.)  

The Fourier spectra plotted in \fref{fig:kineticAndEnstrophySpectra} further illustrate the growth of the low-order SASI modes, the development of turbulence, and subsequent SASI saturation.  
The compensated spectral specific kinetic energy density (i.e., $\widehat{e}_{\vect{u}}\times k^{5/3}$) is plotted versus wavenumber $k$ for select time states in the left panel in \fref{fig:kineticAndEnstrophySpectra}.  
(Enstrophy spectra $\widehat{e}_{\vect{\omega}}$ for the same time states are plotted in the right panel in \fref{fig:kineticAndEnstrophySpectra}.)  
The figure shows the time evolution of the spectra from the early, linear phase of the SASI ($t=600$~ms; cf. left panel in \fref{fig:entropy}), the transition to the nonlinear phase, and well into the nonlinear, saturated state.  

Our simulations suggest that the SASI saturates in the nonlinear regime due to the development of post-shock turbulence.  
In terms of spherical harmonics, the SASI is often characterized by exponentially growing power in low-order modes (e.g., $\ell=1,m=0$) \cite{blondinMezzacappa_2006}.  
As a result, the shock surface deviates rapidly from its initially spherical shape.  
The obliquity of the shock front relative to the pre-shock accretion flow causes the non-radial post-shock kinetic energy to grow at the expense of thermal energy \cite{blondin_etal_2003}.  
This is consistent with our simulations.  
The growth of the low-order SASI mode is clearly seen in the left panel in \fref{fig:kineticAndEnstrophySpectra} (cf. the evolution of the peak around $k=0.03$~km$^{-1}$ from $t=700$~ms to $t=900$~ms).  
Growth of the spectral specific kinetic energy density for larger $k$ values accompanies the growth of the peak, which reaches a maximum around $t=900$~ms.  
The peak in the spectral density is significantly reduced by $t=1000$~ms, when the system has reached an approximate statistically steady state, and the compensated spectral density is nearly flat over a range of scales (i.e., a Kolmogorov-like spectrum $\widehat{e}_{\vect{u}}\propto k^{-5/3}$ is established for $k\in[0.03,0.2]$~km$^{-1}$).  
The dashed line represents the time averaged spectrum, which shows a hint of the peak around $k=0.03$~km$^{-1}$.  
In this saturated state, flows associated with low-order SASI-modes are continuously powered by pre-shock accretion, but the turbulent kinetic energy cascade via secondary fluid instabilities efficiently saps energy from the low-order modes and prevent them from growing further.  
The turbulent energy is eventually converted into heat via viscous dissipation on small spatial scales (or into magnetic energy via a turbulent dynamo, and then converted into heat via Joule dissipation\footnote{In our simulations, viscous and Joule dissipation are due to the shock-capturing scheme adopted \cite{endeve_etal_2012a}}; see \sref{sec:magneticFieldAmplification}).  
Dissipation is clearly dominant for $k\gtrsim0.3$~km$^{-1}$, where the spectral slope of $\widehat{e}_{\vect{u}}$ begins to steepen.  
(Also, the enstrophy spectrum turns over around $k=0.3$~km$^{-1}$.)  

A complementary perspective on the development of post-shock turbulence is gained from the evolution of the enstrophy spectrum in the right panel in \fref{fig:kineticAndEnstrophySpectra}.  
(Enstrophy is proportional to the square of the vorticity magnitude $\Omega=\omega^{2}/2$.)  
Post-shock vorticity is generated baroclinically---and further amplified by vortex stretching---in the strong velocity shear layer separating the counterrotating flows induced by the spiral SASI mode \cite{endeve_etal_2012a}.  
The vorticity spreads quickly throughout the post-shock volume, and is characterized by high spatial and temporal intermittency.  
(\Fref{fig:magneticFieldStrength} shows the spatial distribution of $|\vect{B}|$, which is very similar to that of $|\vect{\omega}|$ \cite{batchelor_1950}.)  
For $t=786$~ms, a growing maximum has formed in the enstrophy spectrum, which peaks on small spatial scales.  
The average wavenumber $\bar{k}_{\omega}=\int_{k_{\mbox{\tiny min}}}^{k_{\mbox{\tiny max}}}k\widehat{e}_{\vect{\omega}}\,dk/\int_{k_{\mbox{\tiny min}}}^{k_{\mbox{\tiny max}}}\widehat{e}_{\vect{\omega}}\,dk$ (marked with a diamond on the three most evolved time states in the right panel in \fref{fig:kineticAndEnstrophySpectra}) is about $0.34$~km$^{-1}$ ($\bar{\lambda}_{\omega}=2\pi/\bar{k}_{\omega}\approx18$~km).  
For later times, the maximum in the enstrophy spectrum continues to increase until $t\approx1000$~ms, while the average wavenumber $\bar{k}_{\omega}$ remains relatively unchanged (it appears to increase slowly with time).  
We suspect that the location of the maximum in the enstrophy spectrum is sensitive to numerical resolution and that $\bar{k}_{\omega}$ will increase with decreasing $\Delta l$ (as is the case with the magnetic energy spectrum \cite{endeve_etal_2012a}) until some for now unspecified physical (\emph{not} numerical) mechanism sets in and prevents $\bar{k}_{\omega}$ from increasing further.  
Nevertheless, in our 3D simulations, the spatial scale associated with the low-order SASI modes ($\sim200$~km) is clearly separated from the spatial scale associated with the turbulence ($\bar{k}_{\omega}\sim20$~km), and turbulence appears to have a singularly destructive effect on the SASI by draining energy from its low-order modes and thereby causing it to saturate.  

\section{Magnetic field amplification from SASI-driven turbulence}
\label{sec:magneticFieldAmplification}

SASI-driven turbulence results in post-shock magnetic field amplification due to an efficient small-scale dynamo \cite{endeve_etal_2010,endeve_etal_2012a}.  
Here we summarize the main characteristics of magnetic field amplification in our simulations.  
Astrophysical turbulence dynamos are broadly divided into small-scale and large-scale dynamos.  
Large-scale dynamos (e.g., the $\alpha\Omega$ dynamo) produce large scale magnetic fields (i.e., the magnetic fields exhibit spatial coherence on scales larger than the flows producing them), and require a significant amount of net kinetic helicity ($\vect{u}\cdot\vect{\omega}$; typically present in shear flows) in order to operate (e.g., \cite{meneguzzi_etal_1981,brandenburg_2001}).  
We find that the turbulence induced by the spiral SASI mode is essentially non-helical, and results in a small-scale dynamo (this characteristic may, however, change if a rapidly and differentially rotating PNS is included in the model).  
Small-scale dynamos result in magnetic fields with a characteristic spatial scale similar to the turbulent flows that produce them.  
(See for example Brandenburg \& Subramanian \cite{brandenburgSubramanian_2005} for a recent review of astrophysical dynamo theory.)  

\begin{figure}
  \begin{center}
    \includegraphics[angle=00,height=3.0in,width=3.0in]{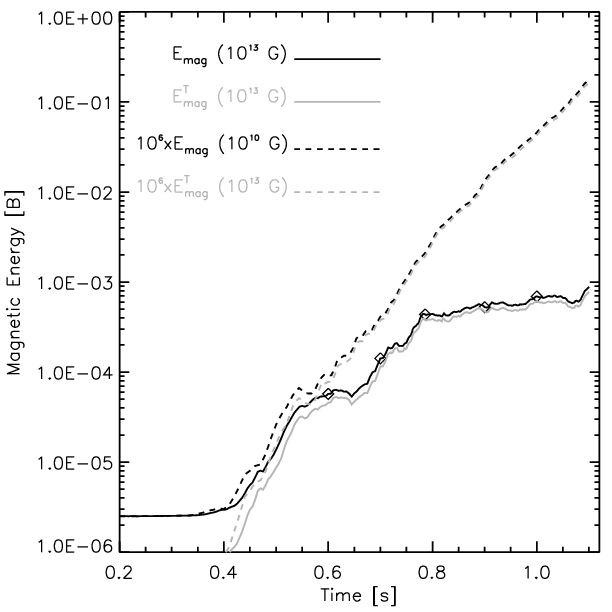}
    \includegraphics[angle=00,height=3.0in,width=3.0in]{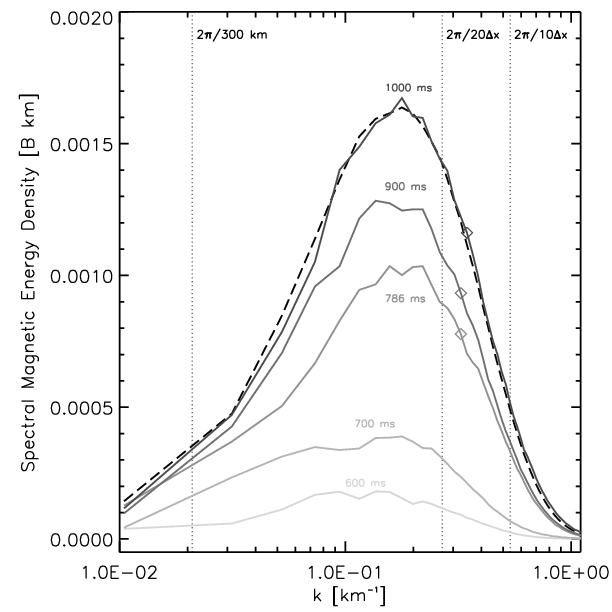}
    \caption{Magnetic field evolution from SASI-driven turbulence.  {\it Left panel:} total magnetic energy $E_{\mbox{\tiny mag}}=\f{1}{2\mu_{0}}\int_{V_{\mbox{\tiny Sh}}}B^{2}\,dV$ (black lines) and ``turbulent" magnetic energy $E_{\mbox{\tiny mag}}^{\mbox{\tiny T}}=\int_{k^{\mbox{\tiny T}}}^{\infty}\widehat{e}_{\mbox{\tiny mag}}(k)\,dk$ ($k^{\mbox{\tiny T}}=0.1$~km$^{-1}$; grey lines) versus time for the weak-field model (dashed lines) and the strong-field model (solid lines).  The results from the weak-field model have been multiplied by $10^{6}$ for easy comparison of the models.  {\it Right panel:} spectral magnetic energy density $\widehat{e}_{\mbox{\tiny mag}}$ at select times during the evolution of the strong-field model.  The diamonds on the solid black line in the left panel correspond to the times (solid lines) displayed in the right panel.  The diamond on the three most advanced time states in the right panel indicates the mean magnetic wavenumber $\bar{k}_{\mbox{\tiny mag}}$ (see text for definition).  The dashed line in the right panel represents the time averaged spectral magnetic energy density $\timeAverage{\widehat{e}_{\mbox{\tiny mag}}(k)}{1.0}{1.1}$.  \label{fig:magneticEnergyAndSpectrum}}
  \end{center}
\end{figure}

The time evolution of the total magnetic energy below the shock is shown in the left panel in \fref{fig:magneticEnergyAndSpectrum} (black lines), for the weak-field model (dashed) and the strong-field model (solid).  
Both models exhibit exponential magnetic energy growth, with growth time $\tau\sim60$~ms, during the early nonlinear evolution of the SASI ($t\lesssim786$~ms).  
The rapid magnetic energy growth in the strong-field model effectively stops around $t=768$~ms, when the magnetic energy density becomes comparable to the kinetic energy density ($B\sim\sqrt{\mu_{0}\rho}u$) in localized regions below the shock.  
The magnetic energy in the strong-field model is about $5\times10^{-4}$~B when it saturates.  
The magnetic energy in the weak-field model grows exponentially with a nearly unchanged growth rate throughout the run, and increases by almost five orders of magnitude.  

The evolution of the magnetic energy spectrum is similar to that of the enstrophy spectrum (\fref{fig:kineticAndEnstrophySpectra}).  
In the right panel in \fref{fig:magneticEnergyAndSpectrum}, the spectral magnetic energy density from the strong-field model is plotted versus wavenumber for select times during the simulation (the same times displayed in \fref{fig:kineticAndEnstrophySpectra}).  
A maximum in the spectrum develops on small spatial scales (around $k=0.2$~km$^{-1}$), which is due to the developing turbulence.  
The average magnetic wavenumber $\bar{k}_{\mbox{\tiny mag}}=\int_{k_{\mbox{\tiny min}}}^{k_{\mbox{\tiny max}}}k\widehat{e}_{\mbox{\tiny mag}}\,dk/\int_{k_{\mbox{\tiny min}}}^{k_{\mbox{\tiny max}}}\widehat{e}_{\mbox{\tiny mag}}\,dk$ (marked with a diamond on the three most evolved time states in the right panel in \fref{fig:magneticEnergyAndSpectrum}) is about $0.32$~km$^{-1}$ (i.e., similar to $\bar{k}_{\omega}$; $\bar{\lambda}_{\mbox{\tiny mag}}=2\pi/\bar{k}_{\mbox{\tiny mag}}\approx20$~km) for $t=786$~ms, and does not change much for later times.  
The maximum in the magnetic energy spectrum continues to increase until $t\approx1000$~ms, when the spectrum is very similar to the time averaged spectrum $\timeAverage{\widehat{e}_{\mbox{\tiny mag}}}{1.0}{1.1}$.  

\begin{figure}
  \begin{center}
    \includegraphics[angle=00,height=4.0in,width=4.0in]{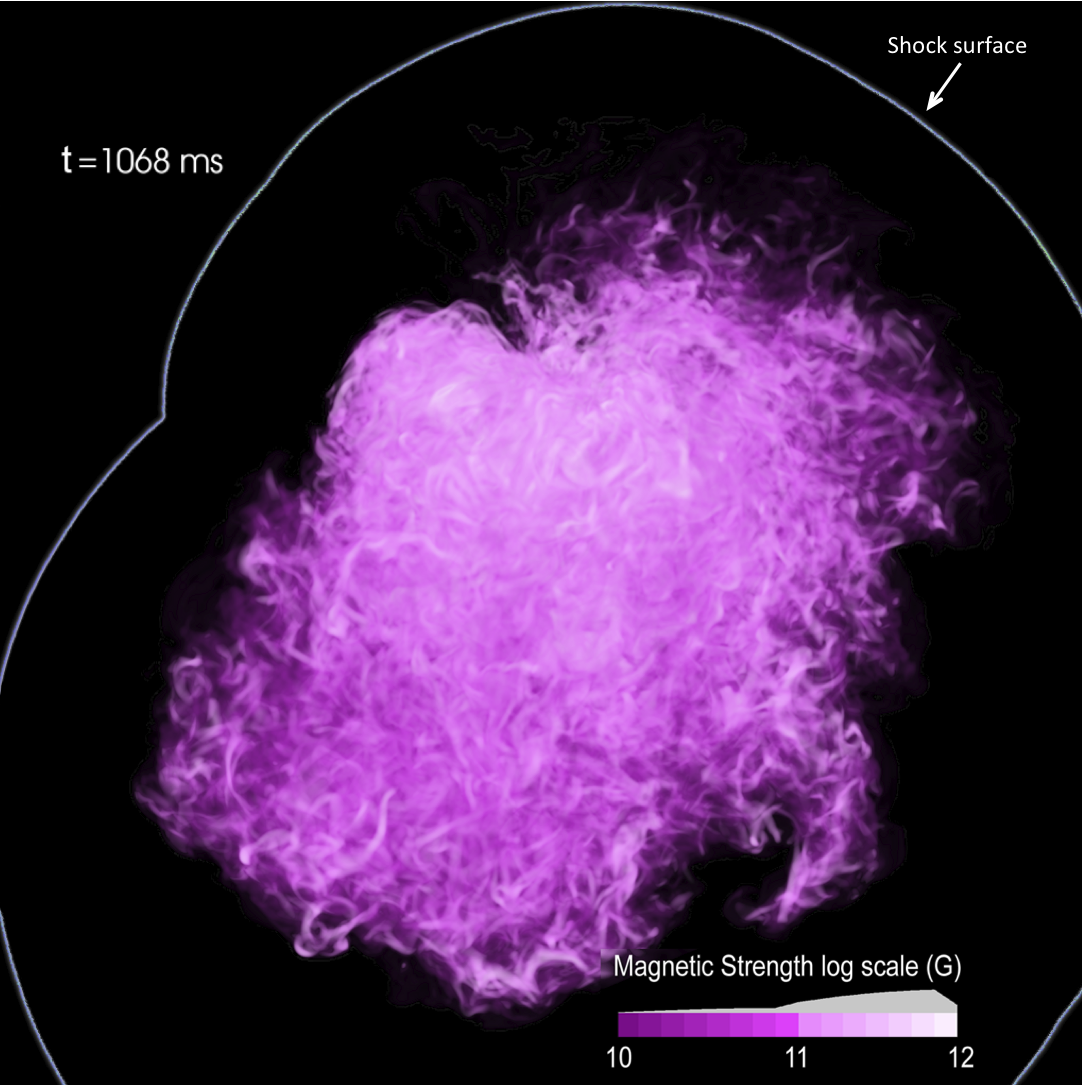}
    \caption{Spatial distribution of the magnitude of the magnetic field during the late, turbulent phase of the weak-field model.  
    Image courtesy of Ross Toedte, ORNL.  \label{fig:magneticFieldStrength}}
  \end{center}
\end{figure}

The magnetic fields induced by SASI-driven turbulence are characterized by a highly intermittent flux tube structure, where all the magnetic energy resides on small spatial scales ($k>0.1$~km$^{-1}$).  
From the grey lines in the left panel in \fref{fig:magneticEnergyAndSpectrum} we see that the growth of the magnetic energy below the shock is due to small-scale fields (i.e., $E_{\mbox{\tiny mag}}\approx E_{\mbox{\tiny mag}}^{\mbox{\tiny T}}$).  
The spatial distribution of the magnetic field magnitude, shown in \fref{fig:magneticFieldStrength} from a late stage in the weak-field model, confirms the turbulent nature of the $B$-fields.  
The net magnetic flux below the shock is essentially zero.  
In \fref{fig:magneticFieldStrength}, we also see that the magnetic fields resides in a volume well below the shock surface.  
(In addition to exhibiting similar spectral evolution, the spatial distribution of the magnetic field and the vorticity field are very similar in our simulations \cite{endeve_etal_2010}.)  
Moreover, the high intermittency (quantified by the kurtosis of the probability density function; e.g., \cite{brandenburg_etal_1996}) of the magnetic and vorticity fields are very similar, and $\vect{B}$ and $\vect{\omega}$ also tend to be aligned or antialigned.  
The observed similarities between vorticity and magnetic field during the fully developed turbulent state are consistent with the predictions by Batchelor \cite{batchelor_1950}.  
Also, the SASI-driven turbulence is similar in many ways to the convectively driven (non-helical) MHD turbulence reported by Brandenburg et al. \cite{brandenburg_etal_1996}.  

In our simulations, magnetic field amplification is due to stretching of magnetic flux tubes by the turbulent flows \cite{endeve_etal_2010}.  
The magnetic flux tubes are ``frozen" in the fluid (i.e., fluid elements remain on the same field line), and, since initially adjacent fluid elements separate exponentially in turbulence (e.g., \cite{ott_1998}), the magnetic field strength grows rapidly in proportion to the stretching (i.e., the increased relative separation of fluid elements on the field line).  
At the same time, the flux tubes undergo a decrease in the scale perpendicular to the stretching.  
In general, the field amplification continues until (i) the field becomes strong enough to resist further stretching by the fluid, (ii) the flux tube cross section becomes so small that further field amplification is prevented by resistive dissipation, or, as is the case for our strong-field model, a combination of (i) and (ii).  

The turbulent magnetic field amplification is affected by finite numerical resolution.  
(We see from the right panel in \fref{fig:magneticEnergyAndSpectrum} that the characteristic wavenumber $\bar{k}_{\mbox{\tiny mag}}$ lies in the diffusive regime of the spectrum.)  
Both the growth rate and saturation amplitude are underestimated by the numerical simulations.  
We can, however, estimate the ``true" growth rate indirectly.  
By adopting a non-ideal electric field with resistivity $\eta$, $\vect{E}=-(\vect{u}\times\vect{B})+\eta\curl{\vect{B}}/\mu_{0}$, Faraday's law gives an approximate growth rate for the magnetic field (e.g., \S55 in \cite{landauLifshitz_1960})
\begin{equation}
  \f{1}{B}\pderiv{B}{t}
  \sim\f{u_{\mbox{\tiny rms}}^{\mbox{\tiny T}}}{\lambda^{\mbox{\tiny T}}}
  \left[1-R_{\rm m}^{-1}\left(\f{\lambda^{\mbox{\tiny T}}}{\lambda_{\mbox{\tiny d}}}\right)^{2}\right], 
  \label{eq:growthRate}
\end{equation}
where the magnetic Reynolds number $R_{\rm m}=\mu_{0}\,u_{\mbox{\tiny rms}}^{\mbox{\tiny T}}\,\lambda^{\mbox{\tiny T}}/\eta$ has been defined in terms of the turbulence scale $\lambda^{\mbox{\tiny T}}$ (equal to $\lambda_{\omega}^{\mbox{\tiny T}}$ or $\lambda_{\mbox{\tiny mag}}^{\mbox{\tiny T}}$) and the turbulent rms velocity $u_{\mbox{\tiny rms}}^{\mbox{\tiny T}}$.  
The dissipation scale is denoted with $\lambda_{\mbox{\tiny d}}$ ($\approx\lambda^{\mbox{\tiny T}}$ in our simulations).  
\Eref{eq:growthRate} states that the magnetic field is amplified if the second term in the square bracket is less than unity.  
Moreover, as long as the magnetic field is weak ($u_{\mbox{\tiny rms}}^{\mbox{\tiny T}}$ is independent of $B$), the amplification is exponential, and the growth rate tends to the inverse turnover time for large $R_{\rm m}$.  
(The magnetic Reynolds number in the PNS may be as large as $10^{17}$ \cite{thompsonDuncan_1993}.)  
With information extracted from the Fourier spectra, we find $u_{\mbox{\tiny rms}}^{\mbox{\tiny T}}/\lambda^{\mbox{\tiny T}}\sim250$~s$^{-1}$, which suggests an exponential growth time of a few milliseconds.  
Since we measure an exponential growth time of the magnetic energy of about 60~ms, we conclude that the effective magnetic Reynolds number in our simulations is larger than unity (not by very much!), but much smaller than physically realistic values.  
Thus, our simulations are underresolved.  
Simulations with higher spatial resolution can accommodate thinner flux tubes, which results in stronger magnetic fields \cite{endeve_etal_2010}.  We also find that the exponential growth rate in the simulations increases with increasing grid resolution.  

The presence of amplified magnetic fields does not result in noticeable effects on the global shock evolution in our simulations, and the $B$-fields are probably unimportant to supernova explosion dynamics.  
This can be understood as a matter of simple energetics.  
Magnetic energy grows at the expense of kinetic energy, and the kinetic energy content in the post-shock flow during vigorous SASI activity ($\sim5\times10^{-2}$~B) is not enough for magnetic fields to become energetically significant to the explosion ($\sim1$~B).  
Moreover, the turbulent kinetic energy accessed for magnetic field amplification only amounts to a fraction of the total kinetic energy below the shock (\fref{fig:kineticEnergyVersusTime} and \fref{fig:magneticEnergyAndSpectrum}).  
These observations suggest a rather passive role of the magnetic fields on the overall shock dynamics.  
On the other hand, the sensitivity of magnetic field amplification and evolution to numerical resolution prevents us from ruling out completely any effect caused by magnetic fields with importance to the explosion dynamics.  

The magnetic fields amplified by SASI-driven turbulence may contribute nontrivially to PNS magnetization.  
By computing the magnetic energy accumulated in the volume occupied by the PNS due to the Poynting flux through the boundary surface at $r=\rPNS$, we estimate that the PNS magnetic field in the strong-field model due to this process alone exceeds $10^{14}$~G.  
The accumulated magnetic energy ($\gtrsim10^{48}$~erg) meets the energy requirements to power the total flare energy released per SGR \emph{and} the persistent X-ray emission \cite{thompsonDuncan_2001}.  
The weak-field model results in less accumulated energy.  
However, given the issues with finite grid resolution discussed above, it seems plausible that small-scale PNS magnetic fields exceeding $10^{14}$~G can form due to the SASI alone, independent of the initial magnetic field strength.  

\section{Summary and discussion}

Results from three-dimensional MHD simulations of the SASI are presented.  
The simulations are initiated from a configuration that resembles the early stalled shock phase in a core-collapse supernova.  
Although our simulations adopt a simplified physics model that excludes neutrino transport, self-gravity, and the PNS itself, the adopted spatial resolution is currently inaccessible to state-of-the-art supernova models in three spatial dimensions.  
Our simulations provide valuable insight into hydrodynamic and MHD developments, which will likely be relevant to multiphysics simulations of core-collapse supernovae.  
We summarize our simulation results as follows
\begin{itemize}
  \item[1.]{Vigorous turbulence develops via secondary hydrodynamic instabilities (e.g., the Kelvin-Helmholtz instability) in the shear layer separating the counterrotating flows induced by the spiral SASI mode, and we find that the specific kinetic energy spectrum develops Kolmogorov-like scaling (i.e., $\widehat{e}_{\vect{u}}\propto k^{-5/3}$) in the wavenumber range separating the post-shock forcing scale (given by the low-order SASI modes) and the dissipation scale (set by finite grid resolution in our simulations).}
  \item[2.]{Our simulations suggest that the SASI saturates nonlinearly due to the development of post-shock turbulence \cite{guilet_etal_2010}.  
  In particular, the evolution of the specific kinetic energy spectrum (\fref{fig:kineticAndEnstrophySpectra}) illustrates how the turbulence saps energy from the low-order SASI modes.  
  In the saturated state, a significant fraction of the post-shock kinetic energy is due to turbulence (with $u_{\mbox{\tiny rms}}^{\mbox{\tiny T}}\sim5\times10^{3}$~km~s$^{-1}$).}
  \item[3.]{The SASI-driven turbulence results in magnetic field amplification by stretching via an efficient small-scale dynamo.  
  The turbulence is essentially non-helical and is similar in many ways to convectively driven MHD turbulence (e.g., \cite{brandenburg_etal_1996}).  
  The magnetic field evolution is sensitive to numerical resolution.  
  In particular, the magnetic energy growth rate and saturation amplitude remain uncertain.  
  However, estimates using data extracted from our simulations suggest that the magnetic energy may grow on a millisecond timescale under more realistic physical conditions (i.e., with $R_{\rm m}\gg1$).}
  \item[4.]{The induced magnetic fields have no noticeable effect on the global shock dynamics, and this can be understood from simple considerations of the energetics:  a relatively small fraction of the post-shock kinetic energy is accessed for magnetic field amplification.  
  On the other hand, SASI-driven turbulence may contribute nontrivially to proto-neutron star magnetization.  
  (Simple estimates suggest PNS magnetic fields exceeding $10^{14}$~G.)  
  Thus, the formation of strongly magnetized proto-neutron stars may not necessarily be linked uniquely to magnetorotationally-driven supernova explosions (e.g., \cite{leblancWilson_1970}), but may merely result as a by-product of the violent dynamics associated with the explosion of massive stars.}
\end{itemize}

Given that the SASI may play a central role in facilitating neutrino powered explosions, our simulations suggest that turbulence---through its role in the nonlinear saturation of the SASI---may play an important role in core-collapse supernova models as well.  
We have intentionally adopted a simplified physics model in order to carry out our simulations with high spatial resolution.  
They therefore represent an initial contribution towards a complete understanding of supernova dynamics, and the role of turbulence in particular.  
(Recently, Murphy \& Meakin \cite{murphyMeakin_2011} proposed a turbulence model for the core-collapse supernova problem.)  
The role of turbulence in supernova models must be further investigated in the context of multiphysics simulations.  
Fortunately, such simulations---in three spatial dimensions---are now becoming available, although initially with limited spatial resolution \cite{takiwaki_etal_2011}.  

An understanding of the impact of the disparate evolution of turbulent flows in 2D and 3D on core-collapse supernova simulations will be important to establish.  
Current state-of-the-art, multiphysics simulations of neutrino-powered explosions \cite{bruenn_etal_2006,buras_etal_2006,bruenn_etal_2009,marekJanka_2009,suwa_etal_2010,muller_etal_2012}, carried out in two spatial dimensions with axial symmetry imposed, emphasize the importance of the SASI in facilitating the explosion.  
The post-shock flows in these simulations are clearly turbulent due to both neutrino-driven convection and the SASI.  
However, it is known that two-dimensional (inviscid and incompressible) turbulence supports a so-called inverse cascade, where smaller vortices of equal size coalesce to form larger vortices, which leads to increased turbulent energy on spatial scales larger than the turbulent driving scale \cite{kraichnan1967,frischSulem_1984}; i.e., opposite to the turbulent cascade in 3D (cf. \fref{fig:kineticAndEnstrophySpectra}).  
Comparisons between 2D and 3D simulations have been carried out in the context of parametrized models \cite{nordhaus_etal_2010,hanke_etal_2011}.  
In particular, Hanke et al. \cite{hanke_etal_2011} observed diverging asymptotic behavior with increasing grid resolution in their 2D and 3D simulations: their 2D models explode more easily with increasing resolution, while the opposite is true for their 3D models, and they attribute these differences to the different cascading nature of 2D and 3D turbulence.  
Clearly, a better understanding of the role of turbulence in 2D and 3D simulations of the SASI (and how they potentially differ) is important to establish.  
We plan to investigate this issue in detail in a future study.  

Neutrino-driven convection develops in the post-shock flow on a timescale that is generally shorter than that of the SASI (e.g., \cite{buras_etal_2006}).  
It will also be important to better understand the interplay between the turbulence induced by neutrino-driven convection and SASI development---in particular, the influence of pre-existing turbulence on the coherence of the low-order SASI modes.  

We cannot completely dismiss turbulence-induced magnetic fields as unimportant to the explosion dynamics.  
With simple considerations of the energetics, we have argued above (and elsewhere \cite{endeve_etal_2010,endeve_etal_2012a}) that the magnetic fields induced by SASI-driven turbulence have little potential to impact the global shock dynamics.  
The presence of strong magnetic fields (model $\model{13}$ versus model $\model{10}$) results in less turbulent kinetic energy \cite{endeve_etal_2012a}.  
However, given the difficulties associated with finite grid resolution when simulating MHD turbulence in large scale simulations such as those presented here, one must use caution when extrapolating conclusions based on low Reynolds number simulations to systems governed by extremely large Reynolds numbers (e.g., \cite{ishihara_etal_2009}).  
Our simulations merely suggest that MHD turbulence (as opposed to hydrodynamic turbulence) may be the more appropriate description of turbulence in core-collapse supernovae, and a better understanding of the impact of magnetic fields is clearly needed here.  
Moreover, with respect to magnetic field evolution in CCSNe, we have not considered (i) the impact of strong differential rotation near the surface of the PNS (which can excite the MRI; e.g., \cite{akiyama_etal_2003,obergaulinger_etal_2009}), or (ii) the impact of neutrino-driven convection.  
Studying the confluence of (i) and (ii) (which alone can excite an $\alpha\Omega$ dynamo \cite{thompsonDuncan_1993}) with turbulent amplification by the SASI is needed to better understand the role of $B$-fields in CCSNe.  
A combination of multiphysics simulations \emph{and} simulations of simplified models (guided by multiphysics simulations) is needed.  

\ack

This research was supported by the Office of Advanced Scientific Computing Research and the Office of Nuclear Physics, U.S. Department of Energy.  
This research used resources of the Oak Ridge Leadership Computing Facility at the Oak Ridge National Laboratory provided through the INCITE program.

\end{document}